\documentstyle[12pt,epsf,here]{article}

\def \be {\begin{equation}}
\def \e {\end{equation}}
\def \bea {\begin{eqnarray}}
\def \ea {\end{eqnarray}}

\def \no {\nonumber}

\def \sub {\scriptscriptstyle}

\newcommand{\To}[2]{\stackrel{#1}{\hbox to #2 pt{\rightarrowfill}}}

\def \vector#1{\stackrel{\hspace{-0.45em}\longrightarrow}{#1}}

\def\np#1#2#3{{\it Nucl.~Phys.\/}~{\bf B#1} (19#2) #3}
\def\pl#1#2#3{{\it Phys.~Lett.\/}~{\bf B#1} (19#2) #3}

\def\cpc#1#2#3{{\it Comput.~Phys.~Commun.\/}~{\bf #1} (19#2) #3}

\begin{document}  
\vspace*{-2cm}  
\renewcommand{\thefootnote}{\fnsymbol{footnote}}  
\begin{flushright}  
hep-ph/9907235\\
DTP/99/62\\  
July 99 \\  
\end{flushright}  
\vskip 65pt  
\begin{center}  
{\Large \bf Anomalous Quartic Couplings\\[3truemm] in \boldmath
$\nu \bar{\nu} \gamma \gamma$ Production via \boldmath $WW$-Fusion
\\[3truemm] 
at LEP2} \\ 
\vspace{1.2cm} 
{\bf  

W.~James~Stirling${}^{1,2}$\footnote{W.J.Stirling@durham.ac.uk}  and
Anja Werthenbach${}^1$\footnote{Anja.Werthenbach@durham.ac.uk} }\\  
\vspace{10pt}  
{\sf 1) Department of Physics, University of Durham,  
Durham DH1 3LE, U.K.\\  
  
2) Department of Mathematical Sciences, University of Durham,  
Durham DH1 3LE, U.K.}  
  
\vspace{70pt}  
\begin{abstract}
The production of $\nu \bar{\nu} \gamma \gamma$ in high-energy 
$e^+e^-$ collisions offers a window on  anomalous
quartic gauge boson couplings. We investigate the effect of
two possible anomalous couplings on the cross section
for $\nu \bar{\nu} \gamma \gamma$ production via $WW$-fusion
at LEP2 ($\sqrt{s} = 200$~GeV).
\end{abstract}
\end{center}  
\vskip12pt

\setcounter{footnote}{0}  
\renewcommand{\thefootnote}{\arabic{footnote}}  
  
\vfill  
\clearpage  
\setcounter{page}{1}  
\pagestyle{plain} 
\section{Introduction} 

In the Standard Model (SM), the couplings of the gauge bosons and fermions
are tightly constrained by the requirements of gauge symmetry.
In the electroweak sector, for example, this leads to trilinear $VVV$ 
and quartic $VVVV$ interactions
between the gauge bosons $V=\gamma, Z^0, W^\pm$ with completely specified
couplings. Electroweak symmetry breaking via the Higgs mechanism gives
rise
to additional Higgs -- gauge boson interactions, again with specified
couplings.\\

The trilinear and quartic gauge boson couplings probe different 
aspects of the weak interactions. The trilinear couplings directly  test
the
non-Abelian gauge structure, and possible deviations from the SM
forms have been extensively studied
in the literature, see for example \cite{tgvtheory} and references therein. 
Experimental bounds have also been obtained \cite{tgvexpt}. 
In contrast, the quartic couplings 
can be regarded as a more direct window on electroweak symmetry breaking,
in particular to the scalar sector of the theory (see for example
\cite{godfrey}) or, 
more generally, on new physics which couples to electroweak bosons. \\

In this respect it is quite possible that the quartic couplings deviate
from their SM values 
while the triple gauge vertices do not. For example,
if the mechanism for electroweak symmetry breaking does not reveal itself
through 
the discovery of new particles such as the Higgs 
boson, supersymmetric particles or technipions it is  
possible that anomalous quartic couplings could provide the first evidence 
of new physics in  this sector of the electroweak theory \cite{godfrey}.
\\

High-energy colliders provide the natural environment for studying
anomalous quartic couplings. The sensitivity of a given process to anomalous
quartic couplings depends on the relative importance of SM contributions to the anomalous contribution, as we shall see. \\

In this study we shall focus on $e^+e^-$ collisions, and quantify the
dependence of the $WW$-fusion $e^+e^- \to \nu \bar{\nu} \gamma \gamma $ cross section on the anomalous couplings. Here we do not consider contributions from $e^+e^- \to Z \gamma \gamma \to \nu \bar{\nu} \gamma \gamma$ but refer to \cite{anomalous} where $e^+e^- \to Z \gamma \gamma $ is studied in detail. Note that in practice the anomalous contributions arising from $WW$-fusion and those from the resonant $Z$ can be added with no danger of double counting, since the anomalous vertex is $WW\gamma \gamma$ in the former case and $ Z Z \gamma \gamma$ in the latter. We shall consider in particular $\sqrt{s} = 200$~GeV corresponding to LEP2, and comment on the effect of increasing the collider energy. \\

Note that
our primary interest is in the so-called `genuine' anomalous quartic
couplings, i.e. those which give no contribution to the trilinear
vertices. \\

In the following section we review the various types of anomalous quartic
coupling relevant for this analysis
that might be expected in extensions of the SM. In Section~3 we present
numerical 
studies illustrating the impact of the anomalous couplings on the $WW$-fusion $\nu \bar{\nu} \gamma \gamma$ cross sections.
Finally in Section~4 we present our conclusions. 

\section{Anomalous gauge boson couplings}

The lowest dimension operators which lead to genuine quartic couplings 
where at least one photon is involved are of dimension 6 \cite{belanger,anomalous}.\\

The
neutral
and the charged Lagrangians, both giving anomalous contributions
to the $WW\gamma\gamma$ vertex, are

\bea
\label{L0}
{\cal L}_0 &=& - \frac{e^2}{16 \Lambda^2}\, a_0\, F^{\mu \nu} \, F_{\mu
\nu} \vector{W^{\alpha}} \cdot \vector{W_{\alpha}} \no \\
&=&  - \frac{e^2}{16 \Lambda^2}\, a_0\, \big[ - 2 (p_1 \cdot p_2 ) 
( A \cdot A) + 2 (p_1 \cdot A)(p_2 \cdot A)\big] \no \\
&& \hspace{1.5cm} {\sub \times} \big[ 2 ( W^+ \cdot W^-) +  (Z \cdot Z) /
\cos ^2 \theta_w \big]  \quad ,
\ea
\bea
\label{Lc}
{\cal L}_c &=& - \frac{e^2}{16 \Lambda^2}\, a_c\, F^{\mu \alpha} \, F_{\mu
\beta} \vector{W^{\beta}} \cdot \vector{W_{\alpha}} \no \\
&=& - \frac{e^2}{16 \Lambda^2}\, a_c\, \big[- (p_1 \cdot p_2)\, A^{\alpha}
A_{\beta} +(p_1 \cdot A)\, A^{\alpha} p_{2 \beta} \big. \no \\
&& \hspace{1.8cm}\big. \quad \quad + (p_2 \cdot A)\,
p_1^{\alpha} A_{\beta} -(A \cdot A)\, p_1^{\alpha} p_{2 \beta} \big] \no
\\
&& \hspace*{1.5cm} {\sub \times} \big[ W_{\alpha}^- W^{+ \beta} +
W_{\alpha}^+ W^{-
\beta} + Z_{\alpha} Z^{\beta} / {\cos ^2 \theta_w} \big] \ . 
\ea
where $p_1$ and $p_2$ are the photon momenta and 

\bea
\label{wb}
\vector{W _{\mu}}  
 = \left( \begin{array}{c}  \frac{1}{\sqrt{2}} (W_{\mu}^+ + W_{\mu}^-) \\  \frac{i}{\sqrt{2}} ( W_{\mu}^+ - W_{\mu}^-) \\  W_{\mu}^3 - \frac{g^{\prime}}{g} B_{\mu} \end{array} \right)
= \left( \begin{array}{c}  \frac{1}{\sqrt{2}} (
W_{\mu}^+ 
+ W_{\mu}^-) \\  \frac{i}{\sqrt{2}} ( W_{\mu}^+ - W_{\mu}^-) 
\\ \frac{Z_{\mu}}{\cos \theta_w}  \end{array} \right)
\ea

\noindent
with  $g^{\prime}=\frac{e}{\cos \theta_w}$ and $g=\frac{e}{\sin \theta_w}$, originating from the requirement of a custodial $SU(2)$ symmetry
to keep the $\rho$ parameter, $\rho = M_W^2/(M_Z^2 \cos ^2
\theta_w)$, close to its measured SM value of 1. \\

It follows from the Feynman rules  that any
anomalous contribution is {\it linear} in the photon energy 
$E_{\gamma}$. This means that it is the hard tail of the
photon energy distribution that is most affected by the anomalous
contributions, but unfortunately the cross section here is very small.
In the following numerical studies we will impose a lower energy photon
cut of $E_{\gamma}^{\rm min} = 20$~GeV. Similarly, there is also
no anomalous contribution to the initial state photon radiation, and
so the effects are largest for centrally-produced photons. We therefore
impose an additional cut of $\vert \eta_\gamma\vert <
2$.\footnote{Obviously 
in practice these cuts will also be tuned to the detector capabilities.}
\\

Finally, the anomalous parameter $\Lambda$ that appears in all the
above anomalous contributions has to be fixed. 
In practice, $\Lambda$ can only be meaningfully specified in the context of a specific model for the new physics giving rise to the quartic couplings. 
However, in order to make our analysis independent of any such model, we choose to fix $\Lambda$ at a reference
value of $M_W$, following the conventions adopted in the literature. Any other choice of $\Lambda$ (e.g. $\Lambda = 1$~TeV) 
results in a trivial rescaling of the anomalous parameters $a_0$ and $a_c$\footnote{For a more detailed discussion of the parameter $\Lambda$ we refer to \cite{anomalous}.}.

\section{Numerical Studies}

\begin{figure}[H]
\hspace*{0.7cm} \centerline{\epsfysize=15.5cm\epsffile{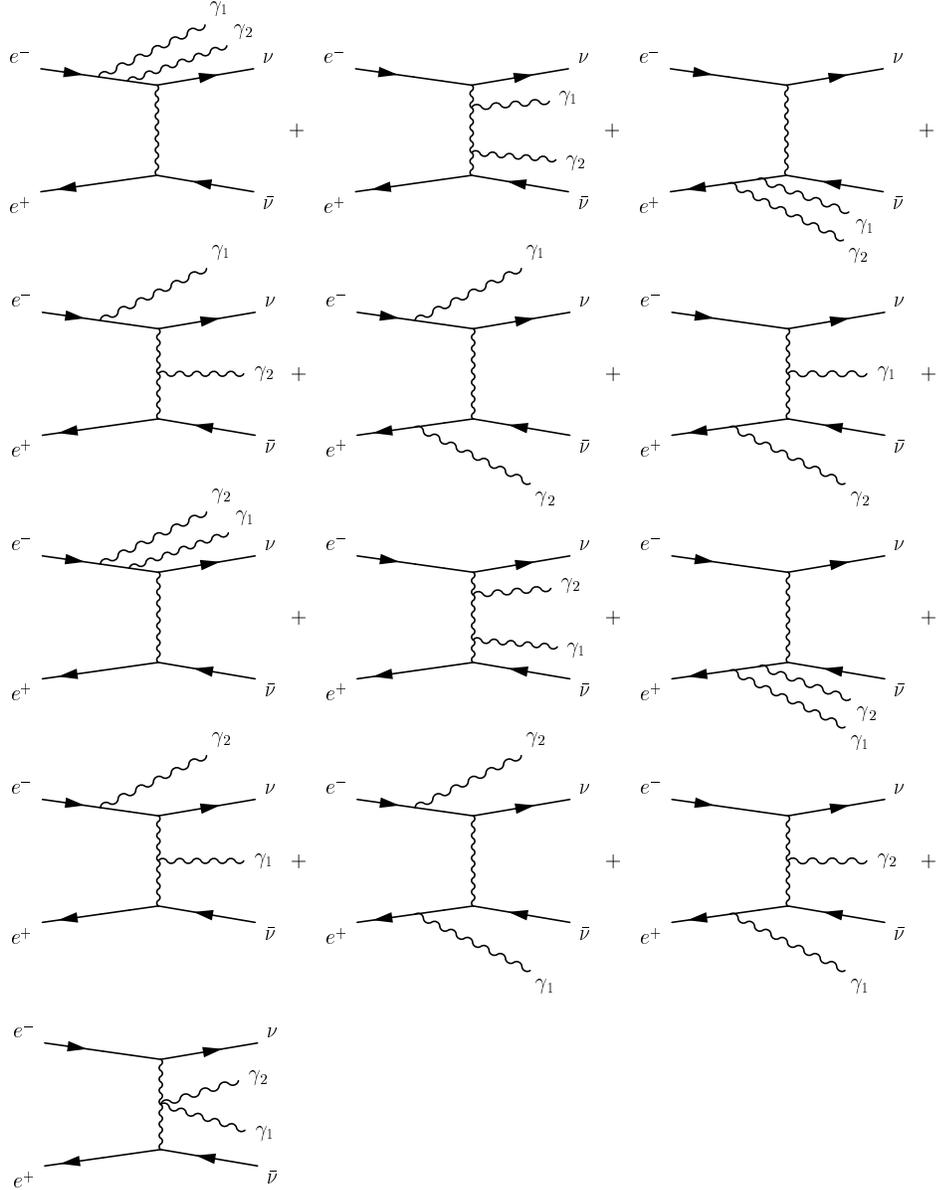}}
\vspace{-0.8cm}
\caption{\label{feyn}{Feynman diagrams contributing to the $WW$-fusion $e^+e^- \to \nu \bar{\nu} \gamma \gamma $ process.}}
\end{figure}

In this section we study the dependence of the $e^+e^- \to \nu \bar{\nu} \gamma \gamma $ $WW$-fusion cross section on the 
two anomalous couplings defined in Section~2. Note that by  '$WW$-fusion` we mean the contribution of the Feynman diagrams shown in Fig.~\ref{feyn} to the cross section.
\vspace*{1cm}

The SM calculation is based on MADGRAPH \cite{madgraph}. 
As already stated, we apply a cut on the photon 
energy $E_{\gamma} > 20$~GeV to take care of the infrared singularity,
 and a cut on the photon rapidity 
 $|\eta_{\gamma}| < 2$ to avoid collinear singularities. \\

As mentioned in the Introduction we do not include contributions from  $e^+e^- \to Z \gamma \gamma \to \nu \bar{\nu} \gamma \gamma$, which obviously do not involve the $WW\gamma\gamma$ vertices. These have been studied in Ref.~\cite{anomalous}\footnote{Note that in Ref.~\cite{anomalous} strictly  $e^+e^- \to Z \gamma \gamma $ has been studied and for comparison with the present $WW$-fusion analysis the branching ratio $\Gamma(Z \to \nu \bar{\nu})$ has to be taken into account as well. This will result in weaker bounds due to the smaller cross section.}. In practice, they can be straightforwardly removed by imposing cuts on the missing mass $M_{\nu \bar{\nu}}\;\, ( M_{\nu \bar{\nu}} < M_Z)$. Nevertheless it has to be said that the  $ZZ\gamma \gamma$ vertex has the identical anomalous structure, only the overall coupling is different. \\

\begin{figure}[H]
\hspace{0.5cm} \centerline{\epsfysize=19cm\epsffile{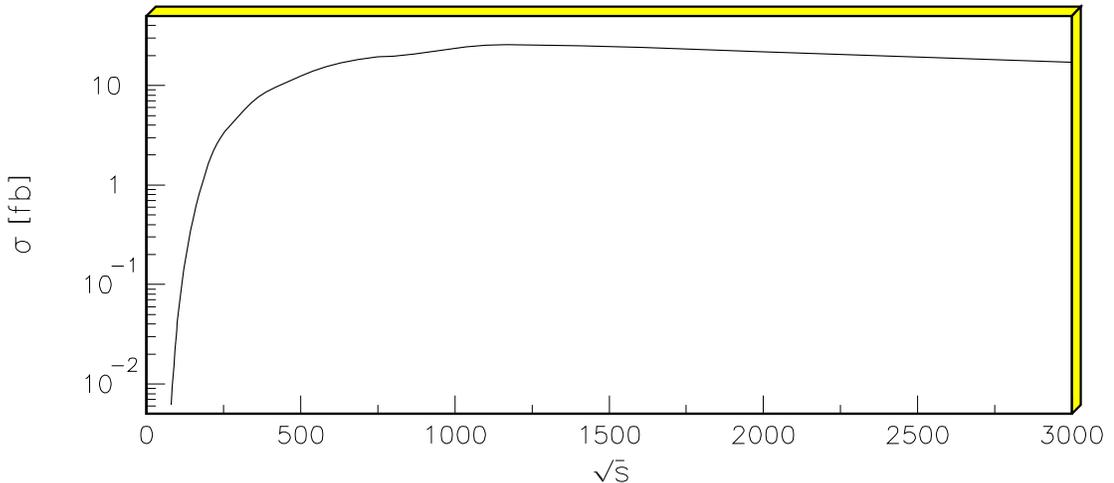}}
\vspace{-10.3cm}
\caption{\label{total}{Total SM cross section for
$e^+e^- \to \nu \bar{\nu} \gamma \gamma$ via $WW$-fusion (in fb)
 as a function of $\protect\sqrt{s}$ with $E_{\gamma} > 20$~GeV and $|\eta_{\gamma}| < 2$.}}
\end{figure}

We first consider the SM cross section for the process of interest,
i.e.
with all anomalous couplings set to zero. Figure~\ref{total} 
shows the collider energy dependence
of the $e^+e^- \to \nu \bar{\nu} \gamma \gamma$ $WW$-fusion cross section.
In the LEP2 energy region the total cross section is {\cal O}(1~fb).
\vspace{-2.5cm}

To study any anomalous effects on the total cross section we need to consider the 
{\it correlations} between the two different anomalous contributions.\\

To obtain quantitative results, 
we consider the experimental scenario of unpolarised $e^+e^- $ collisions
at $200$~GeV with $\int {\cal L} = 150$~pb$^{-1}$.\\

\begin{figure}[H]
\centerline{\epsfysize=17cm\epsffile{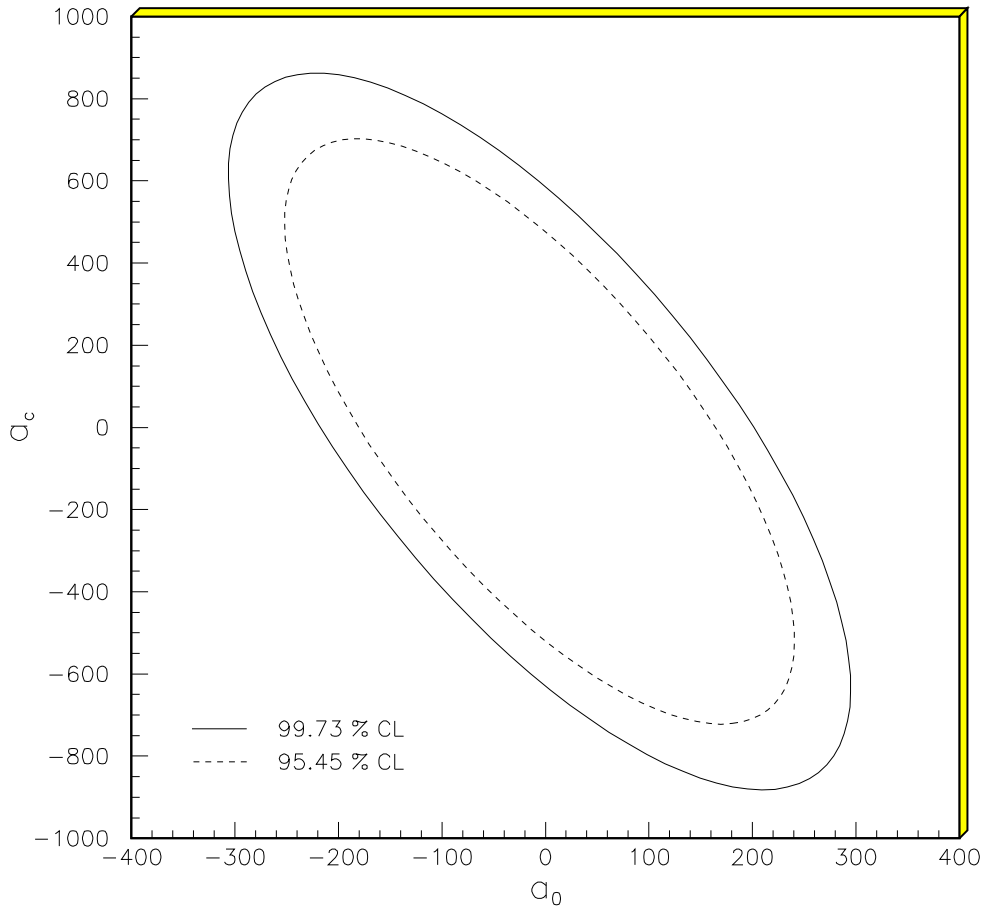}}
\vspace{-2.3cm}
\caption{\label{ellipse}{Contour plots for $+2,+3 \, \sigma$ deviations
from the $WW$-fusion SM $e^+e^-\to \nu \bar{\nu} \gamma \gamma$ total cross section at $\sqrt{s} =
200$~GeV
with $ \int {\cal L} = 150$~pb$^{-1}$.}}
\end{figure}

Figure~\ref{ellipse} shows the contours in the $(a_0,a_c)$ plane that
correspond
to  $+ 2,+3\,\sigma$ deviations from the SM cross section at $\sqrt{s} =
200$~GeV.
\vspace*{-2.0cm}

\section{Discussion and Conclusions} 

We have investigated the sensitivity of the processes $e^+e^- \to
\nu \bar{\nu} \gamma \gamma$  via $WW$-fusion to genuine anomalous
quartic couplings $(a_0,a_c)$ at the canonical centre-of-mass energy
$\sqrt{s}=200$~GeV (LEP2). Key features in
determining the sensitivity  for a given collision energy, apart
from the fundamental process dynamics,  are the
available photon energy $E_{\gamma}$, the ratio of anomalous diagrams to
SM `background' diagrams, and  the polarisation state of the weak bosons
\cite{belanger}. \\

From the purely phenomenological point of view the constraints obtained from this analysis are not competitive with those expected from analysing $WW\gamma$ production and especially from $Z\gamma \gamma$ production. The reason is that although the sensitivity to anomalous contributions is in general increased (i.~e. lower ratio of SM-background to signal and increased phase space due to massless final states) the total cross section itsself is 2 orders of magnitude smaller than those for $WW\gamma$ production or $Z\gamma \gamma$ production. Thus with the relative small luminosity feasible for LEP2 there is little hope that advantages such as the particularly clean experimental environment will make up for the small cross section, and in that case we would expect the tighter bounds on the anomalous parameter to be obtained from analysing $Z\gamma \gamma$ production.  \\

Nevertheless, since only massless particles are produced experimental data from basically any LEP2 centre of mass energy can be used to increase the overall integrated luminosity, and since the process {\it is} highly sensitive to anomalous couplings there is a chance that this process could actually in practice be leading to the tightest bounds. Of course in the end this can only be decided by a proper experimental data analysis.  \\

For a future linear collider with for example $\sqrt{s}=500$~GeV the process $e^+e^- \to
\nu \bar{\nu} \gamma \gamma$ becomes even less competitive, since at that energy the enlarged phase space of massless particles becomes even less important. Note also that at this energy the possibility of producing longitudinally polarised $W,Z$ bosons does increase the sensitivity to anomalous couplings considerably \cite{anomalous}. In the $WW$-fusion process we do not have that opportunity since the $W$s are bound to be `internal' particles with no preferred polarization state.\\

Finally it is important to emphasise that in our study we have only considered `genuine' quartic couplings from new six-dimensional operators. We have assumed that all other anomalous couplings are zero, including the trilinear
ones.  Since the number of possible couplings and correlations 
is so large, it is in practice very difficult to do a combined analysis of {\it all} couplings simultaneously. 

\vspace{0.5cm}
\noindent{\bf Acknowledgements}\\ 

We would like to thank Mark Thomson for drawing our attention to this particular process. We are also grateful to Dave Charlton for fruitful discussions of experimental issues. A.~W. acknowledges the hospitality of the OPAL group.
This work was supported in part by the EU Fourth Framework Programme
`Training and Mobility of 
Researchers', Network `Quantum Chromodynamics and the Deep Structure of
Elementary Particles', 
contract FMRX-CT98-0194 (DG 12 - MIHT). AW gratefully acknowledges
financial support in the form of a
`DAAD Doktorandenstipendium im Rahmen des gemeinsamen Hochschulprogramms
III f\"ur Bund und L\"ander'.\\


\end{document}